# A Scheduler for Unsolicited Grant Service (UGS) in IEEE 802.16e Mobile WiMAX Networks*

Chakchai So-In, *Member, IEEE*, Raj Jain, *Fellow, IEEE*, Abdel-Karim Al Tamimi, *Member, IEEE*

*Abstract*—Most of the IEEE 802.16e Mobile WiMAX scheduling proposals for real-time traffic using Unsolicited Grant Service (UGS) focus on the throughput and the guaranteed latency. The delay variation or delay jitter and the effect of burst overhead have not yet been investigated. This paper introduces a new technique called S̲wapping M̲i̲n-M̲ax (SWIM) for UGS scheduling that not only meets the delay constraint with optimal throughput, but also minimizes the delay jitter and burst overhead.

*Index Terms*—Scheduling; Resource Allocation; Mobile WiMAX; IEEE 802.16e; WiMAX; IEEE 802.16; Unsolicited Grant Service; UGS; Quality of Service; QoS; Delay Jitter

## I. Introduction

One of the key features of the IEEE 802.16e Mobile WiMAX system is its strong quality of service (QoS). IEEE 802.16e Mobile WiMAX provides multiple QoS classes for voice, video, and data applications [1, 2]. To meet QoS requirements especially for voice and video transmissions with delay and delay jitter (delay variation) constraints, the key issue is how to allocate resources among contending users. That is why there are many papers on designing resource allocation algorithms for IEEE 802.16e Mobile WiMAX [2].

The resource in IEEE 802.16e Mobile WiMAX is in units of number of slots. Each slot consists of one subchannel allocated for the duration of some number of OFDM (Orthogonal Frequency Division Multiplexing) symbols. The number of subcarriers in the subchannel and the number of OFDM symbols in the slot depend upon the link direction (uplink, UL, or downlink, DL) and the sub-channelization mode. For example, in the Partially Used Sub-Channelization (PUSC) scheme, one slot consists of one subchannel over two OFDM symbol periods for downlink and one subchannel over three OFDM symbol periods for uplink [2, 3].

Manuscript received February 15, 2010, and revised September 16, 2010. This work was supported in part by the grant from Application Working Group of WiMAX Forum. * "WiMAX," "Mobile WiMAX," "Fixed WiMAX," "WiMAX Forum," "WiMAX Certified," "WiMAX Forum Certified," the WiMAX Forum logo and the WiMAX Forum Certified logo are trademarks of the WiMAX Forum.

C. So-In is with the Department of Computer Science, Faculty of Science, Khon Kaen University, 40002 Thailand (email: chakso@kku.ac.th).

R. Jain and A. Tamimi are with the Department of Computer Science and Engineering, Washington University in St. Louis, One Brookings Drive, Campus Box 1045, Saint Louis, MO, 63130 USA (e-mail: jain and aa7@cse.wustl.edu).

The IEEE 802.16e Mobile WiMAX standard supports bi-directional communication by both frequency division duplexing (FDD) and time division duplexing (TDD). For FDD, the uplink and the downlink use different frequency bands. For TDD, the uplink traffic follows the downlink traffic in time domain. The UGS algorithm discussed in this paper can be used for both FDD and TDD systems. However, to keep the discussion focused, we use a TDD system.

Although the standard allows several configurations such as relay networks, our focus is only on point to multipoint network configuration. Thus, a base station (BS) is the single resource controller for both uplink and downlink directions for all mobile stations (MSs).

IEEE 802.16e Mobile WiMAX offers five classes of service: Unsolicited Grant Service (UGS), extended real-time Polling Service (ertPS), real-time Polling Service (rtPS), non-real-time Polling Service (nrtPS), and Best Effort (BE) classes. UGS is designed for Constant Bit Rate (CBR) traffic with strict throughput, delay, and delay jitter constraints. ertPS is a modification of UGS for voice with silence suppression. rtPS is designed for variable bit rate voice, video, and gaming applications that have delay constraints. nrtPS is for streaming video and data applications that need throughput guarantees but do not have delay constraints (the packets can be buffered). BE is designed for data applications that do not need any throughput or delay guarantees. Note that in practice, the carrier may provide some levels of service guarantee for BE traffic.

These five service classes can be divided in two main categories: non real-time and real-time. nrtPS and BE are in the first category; UGS, rtPS, and ertPS are in the second category. For the first category, common schemes can directly apply such as Weighted Fair Queue (WFQ) and a variation of Round Robin (RR) since there are no hard constraints on delay and delay jitter [2-6]. On the other hand, real-time services have strict constraints on these parameters. This makes scheduling difficult in trying to meet the delay constraint and tolerate the delay jitter with optimal throughput.

UGS is one of the real-time services. Basically, UGS traffic provides a fixed periodic bandwidth allocation. Once the connection is setup, there is no need to send any other requests. UGS is designed and used commonly for Constant Bit Rate (CBR) real-time traffic such as leased-line digital connections (T1/E1) and Voice over IP (VoIP). The main QoS parameters are maximum sustained rate, maximum latency, and tolerated jitter (the maximum delay variation).

As indicated earlier, though there are several proposals [2 to 6] on Mobile WiMAX schedulers, there is not much attention to the effect of delay and delay jitter constraints. In addition, most of the scheduling proposals have ignored the effects of burst overhead. In this paper, we propose an algorithm for UGS scheduling that includes these considerations. Although the discussion in this paper is limited to UGS service class only, we plan to extend this algorithm for other real-time services and for a mixture of users from different service classes.

*Scheduling Factors*

The scheduler for UGS needs to be designed to meet the four main QoS criteria for IEEE 802.16e Mobile WiMAX [1, 2]. First, to optimize system throughput, that is, the scheduler should use all available UGS slots if there is traffic.

Second, the scheduler should guarantee the delay constraints or maximum latency guarantees. In this paper, we also use the term "deadline" to mean delay constraint because the allocation is made within the deadline.

Third, the scheduler should minimize delay jitter. The definition of delay jitter is the variability in inter-packet times from one inter-packet interval to the next.

Finally, the scheduler should minimize number of bursts in order to reduce Media Access Control (MAC) and MAP overheads that reduce system throughput. Note that based on the analysis in [3], the number of bursts has significant impact on the system throughput.

## II. RELATED WORK

There has been some research on delay jitter control for real-time communication in ATM (Asynchronous Transfer Mode) and packet data networks. One way is to introduce a delay jitter regulator or rate regulator at each hop. The regulator delays a whole packet in order to keep constant delay jitter over the end-to-end path [7, 8, 9]. This method minimizes delay jitter with the increase of the mean delay tradeoff.

As shown in our extensive survey of the IEEE 802.16e Mobile WiMAX schedulers [2], channel-unaware IEEE 802.16e schedulers have applied two techniques for UGS traffic: Weighted Round Robin (WRR), Earliest Deadline First (EDF), and equally spread the allocation over all Mobile WiMAX frames (we call this equal allocation or EQA algorithm) [4, 5, 6]. With the admission control, these techniques can achieve optimal throughput and meet deadlines; however, the delay jitter is not considered. This parameter is one of the required QoS parameters for UGS, that is, the tolerated jitter.

In addition, most papers have ignored burst overhead, which directly depends on how many bursts a Base Station (BS) allocates in a Mobile WiMAX frame [3]. Therefore, the delay jitter and the number of bursts are investigated in this paper. In this paper, we introduce a new algorithm, called SWIM (**Sw**apping M**i**n-**M**ax). This algorithm assures deadlines and delay jitter constraints, optimizes the throughput, and also minimizes the number of bursts. In fact, we modified the regulator technique [7, 8, 9] for use in the context of IEEE 802.16e Mobile WiMAX networks. We show that along with zero delay jitter, the number of bursts with SWIM is within a factor of two of those with EDF.

The paper is organized as follows: UGS allocation algorithm with assumptions of arrival traffic and parameters is described in Section III. Then, the SWIM examples are demonstrated in Section IV. Section V shows the performance evaluation. Finally, the conclusions are discussed in Section V.

## III. SWIM ALGORITHM

In this section, general assumptions are described first in Subsection *A* below. Note that our algorithm can be used for both downlink allocation and uplink allocation. However, the problem is more difficult in the uplink since the Base Station (BS) has no information about the actual traffic at Mobile Stations (MSs), e.g., the arrival traffic processes or queue lengths.

The BS only knows about the total demand and the period. Then, in subsection *B*, the SWIM algorithm is introduced. The SWIM algorithm basically can be divided into three basic steps that achieve optimal throughput while meeting the deadline, minimal delay jitter (in fact zero delay jitter), and minimal number of bursts.

### A. Assumptions and Parameter Explanation

Basically for the IEEE 802.16e scheduler, all allocations are integer number of slots. In this paper, the definition of resource is fixed in terms of the number of slots per uplink (or downlink) subframe. This is denoted by the variable *#slots*. The number of bytes corresponding to a slot depends upon the modulation and coding which can vary among users. Without any loss of generality, we use a fixed number of bytes per slot and use bytes as the unit of resource allocation and demand.

For UGS traffic, at connection setup, MSs basically declare the total demand (denoted by *DataSize*) and a *period*. For example, *connection_1* asks 540 bytes every 3 frames. In other words, every 15 ms. Note that WiMAX profiles specify a frame size of 5ms [10]. Due to the periodic nature of UGS traffic, the period is the same as deadline. We use these terms interchangeably; however, we show in Section IV.D that if the deadline is less than the period, the throughput is not optimized.

MSs can dynamically join and leave the networks. For joining, in order for the BS to admit a connection, the BS needs to verify if there are enough resources. Also, the MS may request to change its service on the fly. Therefore, the scheduler needs to be aware of the quality assurance of all currently accepted connections.

In other words, there is an admission control mechanism. The BS can only admit a connection if and only if the sum of the total number of currently used resources per frame and the new demand divided by the deadline of the connection is less than the total available slots per frame for UGS.

The allocation algorithm is based on *DataSize* which is a MAC Service Data Unit (SDU) size for both deadline and delay jitter calculation. We do not explicitly consider any headers such as fragmentation and packing headers, MAC header, and ARQ (Automatic Repeat request) retransmission overhead. However, the MS has to include these overheads in its demand at the connection setup time.

Finally, we assume that the MS has data available at the beginning of each period. In other words, the MS has enough buffer space at least for one period. This allows the BS to allocate the resources anytime within the deadline.

### B. Algorithm Description

Our algorithm has two parts. First, an initialization procedure that starts with optimal throughput and delay. Second, a series of resource swapping steps that leads to optimization of all goals.

A pseudocode showing the nesting of various steps is presented in Fig. 1. Notice that the computational complexity of SWIM in the worst case is in the order of $O(n^2 \log n)$, where $n$ is the number of active connections [11].

```
Preallocation(flows)                              //1st step
Sorted_max_to_min = Sort (flows)
FOR each max_res in Sorted_max_to_min             //2nd step
    Sorted_min_to_max = Sort (flows)
    FOR each min_res in Sorted_min_to_max         //3rd step
        Max_Min_Swapping (max_res, min_res);
    END FOR
END FOR
```

Fig. 1. Steps in SWIM Algorithm

$$Complexity = O(allocations + sorting) = O(n^2 \log n)$$

However, with the known information about the number of flows, the complexity can be reduced to $O(n^2)$. The second sorting process (*Sorted_min_to_max*) is not required; the swapping is processed from the last element of *Sorted_max_to_min* to the beginning.

The steps of SWIM are as follows: First, given $n$ users with $i^{th}$ user demanding data size $d_i$ over a period $p_i$, optimal throughput can be obtained by taking a Least Common Multiple (LCM) of periods $p_i$'s and allocating resources over this cycle.

To achieve zero delay jitter, the algorithm initializes allocated resources (*#slots* or *#bytes*) for each connection by *DataSize/period*, i.e., $d_i/p_i$.

In each frame, the connection with the maximum resource allocation is called *max-res* connection, and the one with the minimum allocation is called *min-res* connection.

Next, to minimize the number of bursts, there is a swapping procedure between *max-res* and *min-res* connections that results in eliminating the *min-res* connection and thereby, reducing the number of connections served in that frame by one. In effect, this reduces the number of bursts in that frame by one. This will become more clear in Section IV, where we provide an example.

The swapping procedure is described as follows: first, the algorithm determines the *min-res* connection, i.e., $i^{th}$ connection and the *max-res* connection, i.e., $j^{th}$ connection. The two connections swap their resources such that $i^{th}$ connection gives up its resources in the current frame while gaining an equal amount of resources in a future frame. Of course, the constraints are that $j^{th}$ connection still needs more resources in this frame, and that $i^{th}$ connection's deadline will still be met.

The system manager can set a minimum burst size parameter, *MinBurstSize*. The swapping procedure ensures that each burst is at least this size. In our examples, we use a *MinBurstSize* of *1*. However, the procedure can be easily applied for any other values of this parameter. The main effect of this parameter is that the connections whose deadline is in the current frame must have *MinBurstSize* allocation or more. If their allocation is equal to *MinBurstSize*, they are excluded from swapping. Leaving *MinBurstSize* at a non-zero value ensures that all SDUs are delivered exactly at the deadline and the delay jitter is zero. Setting *MinBurstSize* to zero will result in a reduced number of bursts but non-zero delay jitter. The SWIM algorithm will then produce results similar to EDF.

The new *max-res* and *min-res* connections do the resource swapping. Note that the total allocated resources per frame do not change by this swapping procedure. Also, the total resources allocated to a connection over its period do not change.

There are a few special cases. First, a *max-res* connection cannot accept more resources than it needs and so the *min-res* connection may not get eliminated. In this case, the next *max-res* connection becomes the candidate for swapping for the remaining resources of the *min-res* connection.

Second, if there are more than one *max-res* connections (more than one connection with the same maximum resources allocated in the frame), we choose the connection whose resources are higher in the next frame.

Third, if there are more than one *max-res* connections with the same next frame resources, we select the connection whose deadline is longer. Of course, we exclude the connections whose deadline is in the current frame and which have allocation equal to *MinBurstSize*.

Fourth, if there are more than one *min-res* connections, we select the connection that has earlier deadline. In case there are more than one *min-res* connections with the same deadline, we choose the connection with lower resources in the next frame.

## IV. SWIM EXAMPLES

In this section, we evaluate the performance of the proposed algorithm with other two commonly used algorithms: EDF and EQA (allocating *DataSize/period*, $d_i/p_i$ in each frame to the $i^{th}$ user). For all three algorithms, the process is cyclic that repeats after LCM period. We show just one such cycle.

First, we evaluate the performance in terms of throughput, mean delay, mean delay jitter, and number of bursts for each algorithm. Then, the concept of flow admission is discussed. Finally, we show an alternative scenario in which the deadline is less than the period. In that case, all resources cannot be allocated to UGS connections optimally. Some resources are left over and can be used by other service classes.

### A. Throughput, Mean Delay, Mean Delay Jitter, and Number of Bursts

The throughput, mean delay, mean delay jitter, and number of bursts are investigated in this section. We start with a simple example (Table I) of static flows by applying all three algorithms: EQA, EDF, and SWIM. The performance comparisons are summarized in Table VI.

Table I shows a simple example of 5 connections (C1 through C5) and their demands (*DataSize*) in bytes and period in terms of WiMAX frames. The total allocated UGS slots are 420 bytes per frame (540/3) + (80/4) + (900/6) + (120/6) + (600/12). With all three algorithms, within one LCM cycle (12 frames in this example), the throughput is optimal, that is, (540×4) + (80×3) + (900×2) + (120×2) + (600×1) = 5,040 bytes or it is equal to 420×12 = 5,040 bytes.

TABLE I
EXAMPLE I: STATIC FLOWS

|  | C1 | C2 | C3 | C4 | C5 |
|---|---|---|---|---|---|
| DataSize (bytes) | 540 | 80 | 900 | 120 | 600 |
| Period (frame) | 3 | 4 | 6 | 6 | 12 |

Tables II and III show the allocations using EQA and EDF algorithms respectively. In EQA, the resource is allocated equally in every frame, e.g., 180 bytes in every frame for C1, 20 bytes for C2, and so on.

TABLE II
EXAMPLE I: EQA ALLOCATIONS

| Time | C1 | C2 | C3 | C4 | C5 | Sum |
|---|---|---|---|---|---|---|
| 0 | 180 | 20 | 150 | 20 | 50 | 420 |
| 1 | 180 | 20 | 150 | 20 | 50 | 420 |
| 2 | 180 | 20 | 150 | 20 | 50 | 420 |
| 3 | 180 | 20 | 150 | 20 | 50 | 420 |
| 4 | 180 | 20 | 150 | 20 | 50 | 420 |
| 5 | 180 | 20 | 150 | 20 | 50 | 420 |

………………….

TABLE III
EXAMPLE I: EDF ALLOCATIONS

| Time | C1 | C2 | C3 | C4 | C5 | Sum |
|---|---|---|---|---|---|---|
| 0 | 420 |  |  |  |  | 420 |
| 1 | 120 | 80 | 220 |  |  | 420 |
| 2 |  |  | 420 |  |  | 420 |
| 3 | 420 |  |  |  |  | 420 |
| 4 | 120 |  | 260 | 40 |  | 420 |
| 5 |  | 80 |  | 80 | 260 | 420 |
| 6 | 420 |  |  |  |  | 420 |
| 7 | 120 |  | 300 |  |  | 420 |
| 8 |  |  | 420 |  |  | 420 |
| 9 | 420 |  |  |  |  | 420 |
| 10 |  |  | 80 |  | 340 | 420 |
| 11 | 120 | 80 | 100 | 120 |  | 420 |

In EDF, the resource is allocated to the connection whose deadline is earliest. At the beginning, C1 has the earliest deadline, that is, 3 frames. In the first frame, the EDF scheduler allocates the entire available capacity of 420 bytes to C1. In the next frame, the scheduler allocates the remaining 120 bytes for C1 to meet C1's throughput guarantee (540 bytes). Of the left-over 300 bytes, 80 and 220 bytes are allocated for C3 and C2, respectively, because the deadlines of C3 and C2 are 4 and 6 frames.

In SWIM, we initialize the allocation table with equal allocation. This results in allocations shown in Table II for EQA. The swapping steps of SWIM are shown in Table IV. In the first frame, the *max-res* connection is C1 and the *min-res* connection is C2. Therefore, C2's allocation in the frame is given to C1 and taken back in the second frame. This results in C1 obtaining 180+20=200 and C2 obtaining 20-20=0 in the first frame. C1 obtains 180-20=160 and C2 obtains 20+20=40 in the second frame. The resulting allocations are shown in Table IV(a). Thus, swapping has reduced the number of bursts by one (one less burst in the first frame while still meeting all the throughput and delay guarantees for all sources).

In the next swapping step, C1 and C4 swap their allocations in frame 1 and 2 resulting in allocations shown in Table IV(b). Next, C1 and C5 swap their allocations in frame 1 and 2 resulting in allocations shown in Table IV(c). Next C1 and C3 swap in frames 1 and 2. However, in this case, C1 has only 90 units of allocations in the second frame and so the swap is done in two steps.

In the first step, 90 units are swapped between C1 and C5 in frames 1 and 2. Then, the remaining 50 units are swapped in frames 1 and 3. This results in allocations shown in Table IV(d). At this point, the allocation for the first frame is complete since there is only one burst left in this frame. Continuing these processes for the second frame and other subsequent frames result in the final allocations shown in Table V.

TABLE IV
EXAMPLE I: EXAMPLE I: SWIM INITIAL STEPS

(a)

| Time | C1 | C2 | C3 | C4 | C5 | Sum |
|---|---|---|---|---|---|---|
| 0 | **200** |  | 150 | 20 | 50 | 420 |
| 1 | **160** | 40 | 150 | 20 | 50 | 420 |
| 2 | 180 | 20 | 150 | 20 | 50 | 420 |

………………….

(b)

| Time | C1 | C2 | C3 | C4 | C5 | Sum |
|---|---|---|---|---|---|---|
| 0 | **220** |  | 150 |  | 50 | 420 |
| 1 | **140** | 40 | 150 | **40** | 50 | 420 |
| 2 | 180 | 20 | 150 | 20 | 50 | 420 |

………………….

(c)

| Time | C1 | C2 | C3 | C4 | C5 | Sum |
|---|---|---|---|---|---|---|
| 0 | **270** |  | 150 |  |  | 420 |
| 1 | **90** | 40 | 150 | **40** | **100** | 420 |
| 2 | 180 | 20 | 150 | 20 | 50 | 420 |

………………….

(d)

| Time | C1 | C2 | C3 | C4 | C5 | Sum |
|---|---|---|---|---|---|---|
| 0 | **420** |  |  |  |  | 420 |
| 1 |  | 40 | **240** | **40** | **100** | 420 |
| 2 | **120** | 20 | **210** | 20 | 50 | 420 |

………………….

TABLE V
EXAMPLE I: FINAL ALLOCATIONS OF SWIM

| Time | C1 | C2 | C3 | C4 | C5 | Sum |
|---|---|---|---|---|---|---|
| 0 | 420 |  |  |  |  | 420 |
| 1 |  |  | 420 |  |  | 420 |
| 2 | 120 | 20 |  |  | 280 | 420 |
| 3 | 360 | 60 |  |  |  | 420 |
| 4 |  |  | 420 |  |  | 420 |
| 5 | 180 | 60 | 60 | 120 |  | 420 |
| 6 | 420 |  |  |  |  | 420 |
| 7 |  | 20 | 400 |  |  | 420 |
| 8 | 120 |  |  |  | 300 | 420 |
| 9 | 420 |  |  |  |  | 420 |
| 10 |  |  | 420 |  |  | 420 |
| 11 | 120 | 80 | 80 | 120 | 20 | 420 |

The mean delays for both EQA and SWIM are the same. These delays are equals to the periods: 3, 4, 6, 6, and 12 frames for connection 1 through 5, respectively. For EDF, the mean delays are $\{(2+2+2+3)/4\}=9/4$, $\{(2+2+4)/3\}=8/3$, $\{(5+6)/2\}=11/2$, $\{(6+6)/2\}=6$, and 11 frames for connections 1 through 5, respectively.

Both EQA and SWIM have zero mean delay jitter, i.e., all SDUs are received on the period. For EDF, the mean delay jitters are $\{(0+0+1)/3\}=1/3$, $\{(0+2)/2\}=1$, 1, 0, and 0 for connections 1 through 5, respectively.

Consider the number of bursts: EQA gives 5 connections × 12 frames or 60 bursts, 24 bursts for SWIM, and 23 bursts for EDF. All four performance metrics are summarized comparatively in Table VI.

TABLE VI
PERFORMANCE COMPARISONS OF UGS SCHEDULING DISCIPLINES

|  | Mean Delay | Mean Delay Jitter | #Bursts | Throughput |
|---|---|---|---|---|
| EQA | Period | Zero | High | Optimal |
| EDF | Low | **Variable** | Low | Optimal |
| SWIM | Period | Zero | Low | Optimal |

*B. Fractional Resource Demand*

Since both EQA's final allocation and SWIM's initial allocation are obtained by dividing the resource demand by the period, this can result in fractional allocations. To show this, we change the resource demands of C1 and C4 in the previous example to 500 and 200, respectively. With a period of 3, C1 requires (500/3) bytes per frame. Similarly, C4 needs 200/6 bytes per frame. With fractional allocations, we simply round the allocations in a frame after the frame has been completely allocated. We find that two decimal digit representations ($1/100^{th}$) are generally sufficient to avoid any truncation errors. Table VII shows the final SWIM allocations for the example. The allocation is still feasible and results in 26 bursts.

TABLE VII
EXAMPLE II: SWIM WITH DATASIZE/PERIOD IS PRIME

| Time | C1 | C2 | C3 | C4 | C5 | Sum |
|---|---|---|---|---|---|---|
| 0 | 420 |  |  |  |  | 420 |
| 1 |  |  | 420 |  |  | 420 |
| 2 | 80 | 7 |  | 33 | 300 | 420 |
| 3 |  | 73 | 347 |  |  | 420 |
| 4 | 420 |  |  |  |  | 420 |
| 5 | 80 | 40 | 133 | 167 |  | 420 |
| 6 | 420 |  |  |  |  | 420 |
| 7 |  | 40 | 380 |  |  | 420 |
| 8 | 80 |  |  | 41 | 299 | 420 |
| 9 |  |  | 420 |  |  | 420 |
| 10 | 420 |  |  |  |  | 420 |
| 11 | 80 | 80 | 100 | 159 | 1 | 420 |

*C. Dynamic Connections*

It is common for flows to join or leave the network. For EQA, a newly admitted flow does not affect the current flows as long as the sum of the total pre-allocated resources and the resource demand per frame of the new connection is less than the total available resources per frame.

The above statement also holds for SWIM and EDF. However, the scheduler needs to maintain the flow states such as how many resources have already been allocated to each connection. We illustrate this with an example. Suppose a new connection C6 joins the network at time 15 with a resource demand of 500 bytes over a period of 4 frames. At $15^{th}$ frame, the total resource demand changes from 420 to 545 bytes per frame. Table VIII shows the initial allocation process for SWIM. The allocations from time 0 to 10 are the same as that in Section IV.A, Table V.

At the end of $14^{th}$ frame, the allocations for the five connections are 540, 20, 420, 0, and 280 bytes. Also, a connection C2 has an allocation of 60 bytes in $15^{th}$ frame. This was a result of previous swapping. In Table VIII, this type of pre-allocation (which has changed from initial value due to swapping) is indicated by enclosing it in parentheses. The pre-allocations for other connections and other frames at the end of $14^{th}$ frame are also shown in the table by enclosing the allocations in parentheses.

To meet their throughput guarantees, in their period containing $15^{th}$ frame, connections C1 through C5 need 540, 40, 300, 0, and 0 bytes over and above their pre-allocations. So, the new initial allocations are made by equally dividing these remaining values by the remaining period. The final results after C6 joins in $15^{th}$ frame are shown in Table IX.

TABLE VIII
EXAMPLE III: SWIM INITIAL STEPS

(a)

| Time | C1 | C2 | C3 | C4 | C5 | **C6** | Sum |
|---|---|---|---|---|---|---|---|
| ….. |  |  |  |  |  |  | 420 |
| 11 | 120 | 80 | 80 | 120 | 20 |  | 420 |
| 12 | 420 |  |  |  |  |  | 420 |
| 13 |  |  | 420 |  |  |  | 420 |
| 14 | 120 | 20 |  |  | 280 |  | 420 |
| 15 | 180 | (60) | (180) | (0) | (0) | 125 | 545 |
| 16 | 180 | 20 | 150 | (70) | (0) | 125 | 545 |
| 17 | 180 | 20 | 150 | (50) | (20) | 125 | 545 |

(b)

| Time | C1 | C2 | C3 | C4 | C5 | **C6** | Sum |
|---|---|---|---|---|---|---|---|
| 15 | **305** | 60 | 180 | 0 | 0 |  | 545 |
| 16 | **55** | 20 | 150 | 70 | 0 | **250** | 545 |
| 17 | 180 | 20 | 150 | 50 | 20 | 125 | 545 |

(c)

| Time | C1 | C2 | C3 | C4 | C5 | **C6** | Sum |
|---|---|---|---|---|---|---|---|
| 15 | **485** | 60 |  | 0 | 0 |  | 545 |
| 16 |  | 20 | **205** | 70 | 0 | 250 | 545 |
| 17 | **55** | 20 | **275** | 50 | 20 | 125 | 545 |

TABLE IX
EXAMPLE III: SWIM WITH A NEW ADMITTED FLOW C6

| Time | C1 | C2 | C3 | C4 | C5 | **C6** | Sum |
|---|---|---|---|---|---|---|---|
| ….. |  |  |  |  |  |  | 420 |
| 11 | 120 | 80 | 80 | 120 | 20 |  | 420 |
| 12 | 420 |  |  |  |  |  | 420 |
| 13 |  |  | 420 |  |  |  | 420 |
| 14 | 120 | 20 |  |  | 280 |  | 420 |
| 15 | 485 | 60 |  | 0 | 0 |  | 545 |
| 16 |  |  | 170 |  | 0 | 375 | 545 |
| 17 | 55 | 60 | 310 | 120 |  |  | 545 |
| 18 | 420 |  |  |  |  | 125 | 545 |
| 19 |  | 20 | 525 |  |  |  | 545 |
| 20 | 120 |  |  |  |  | 425 | 545 |
| 21 | 539 |  |  |  | 6 |  | 545 |
| 22 |  |  | 374 |  | 96 | 75 | 545 |
| 23 | 1 | 80 | 1 | 120 | 218 | 125 | 545 |

*D. Deadline less than the Period*

If the deadline is less than the period, the total demand before and after a connection's deadline is different, and it is possible that some frames may be under-allocated. In other words, it is not possible to achieve full throughput. The unallocated resource can easily be used for non-time critical service classes. If we try to achieve full throughput with UGS traffic only, we may not be able to meet the deadline. This is true for all three algorithms as shown by the examples below.

In Example IV shown in Table X, the deadline for connections C1 through C5 has been set to 2, 4, 4, 4, and 6 frames, respectively. If we allocate equal resources over all frames before the period (resource demand divided by the period), the allocation per frame is 420. We have full throughput but are missing the deadlines. If we allocate equal resources over all frames before the deadline (resource demand divided by the deadline), we need $(540/\mathbf{2}) + (80/4) + (900/\mathbf{4}) + (120/\mathbf{4}) + (600/\mathbf{6})$ or 645. This is over the available capacity of 420. Of course, if the admission control ensures that no new connections will be admitted if the sum of resources per frame (using resource/deadline) is more than the available capacity, we have a feasible solution and can meet the deadlines in a straightforward manner.

TABLE X
EXAMPLE IV: DEADLINE < PERIOD

|  | C1 | C2 | C3 | C4 | C5 |
|---|---|---|---|---|---|
| Data Size (bytes) | 540 | 80 | 900 | 120 | 600 |
| Period (frame) | 3 | 4 | 6 | 6 | 12 |
| Deadline (frame) | 2 | 4 | 4 | 4 | 6 |

## V. PERFORMANCE EVALUATION

In this section, we present numerical results in terms of average number of bursts and delay jitter of EQA, EDF, and SWIM algorithms based on the performance evaluation parameters specified in Mobile WiMAX System Evaluation Methodology documents and WiMAX profiles [3, 10, 12]. These parameters are shown in Table XI.

In this analysis, we use OFDMA PHY with 10 MHz system bandwidth, 5 ms frame, 1/8 cyclic prefix, and a DL:UL ratio of 2:1. Note that 1.6 symbol-columns are used for TTG (Transmit to Transmit Gap) and RTG (Receive to Transmit Gap).

TABLE XI
PERFORMANCE EVALUATION PARAMETERS

| Parameters | Values |
|---|---|
| PHY | OFDMA |
| Duplexing Mode | TDD |
| Frame Length | 5 ms |
| System Bandwidth | 10 MHz |
| FFT size | 1024 |
| Cyclic prefix length | 1/8 |
| DL permutation zone | PUSC |
| RTG + TTG | 1.6 symbol |
| DL:UL ratio | 2:1 (29: 18 OFDM symbols) |
| DL Preamble | 1 symbol-column |
| MAC PDU size | Variable length |
| ARQ and packing | Disable |
| Fragmentation | Enable |
| DL-UL MAPs | 4 symbol-columns |

The number of downlink symbol-columns per frame is 29 [3, 10]. Of these 1 symbol-column is used for preamble and 4 symbol-columns for Frame Control Header (FCH), DL MAP, and UL MAP (repetition of 4), leaving 24 symbol-columns for data transmission. We do not include the optional 4 symbol-columns used to transmit Downlink Channel Descriptor (DCD) and Uplink Channel Descriptor (UCD) in this analysis.

In the Partial Usage of Subchannels (PUSC) mode, there are 30 subchannels in the downlink, and each slot consists of one channel over a two symbol duration. As a result, there are $30 \times (24/2) = 360$ downlink slots per frame. We will use this number for per-frame resource allocation. The number of uplink slots per frame can be also derived similarly. For example, suppose 3 symbol-columns are used for uplink frame overhead (ranging, acknowledgement, etc.), that leaves (15/3) = 5 (#tile symbols) × 35 (#uplink subchannels) = 175 slots [3].

### A. Numerical Configurations

To simplify the analysis, there is only one base station and $N$ number of mobile stations. The base station functions as a centralized scheduler. We assume that the scheduler strictly allocates the resource allocation in each frame. In general, the UGS traffic is fixed size over a pre-defined deadline. We assume the traffic is en-queued at the beginning of each MS period.

Although the standard allows multiple bursts per mobile station (MS), we assume that each MS is limited to only one burst. This minimizes the MAP overhead in the frame. The resource allocation for each MS is randomly generated in range from 1 to 360 slots. Notice that this random slot generation represents a variety of wireless channel conditions (different MCSs).

In addition, the maximum delay (period) is randomly generated from 4 to 44 frames. There are 10 mobile stations. The scheduler allocates the resources over 100 frames. The analysis is over 100 trials. The performance metrics are the mean and the standard deviation of number of bursts in each frame.

### B. Numerical Results

Fig. 2 shows the numerical results for the number of bursts in each 5 ms frame. Over 100 trials, the mean and the standard deviation of number of bursts in each frame for EQA, EDF, and SWIM are 10.00, 1.57, and 2.24; and 0.00, 0.89, and 1.52, respectively. For delay jitter, both SWIM and EQA result in zero delay jitter. The mean and the standard deviation of EDF's delay jitter are 1.36 and 1.08 frames.

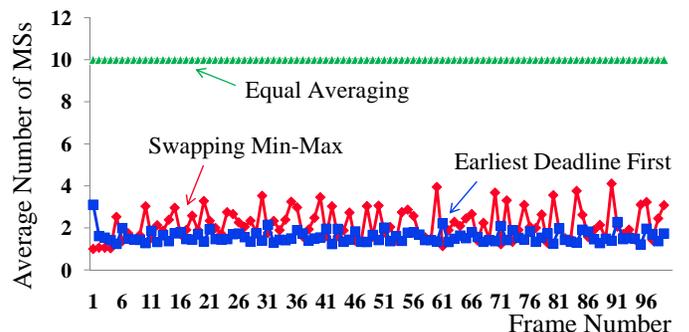

Fig. 2. Number of Bursts for SWIM, EDF, and EQA (10 mobile stations over 100 frames)

We also performed the performance evaluation with more users over longer periods; however, the results are similar to those of the 10-MS scenario.

In general, the results here correspond to the claims stated in Table VI. Again, all three algorithms can achieve the optimal throughput with delay bound (with the admission control). The number of bursts of SWIM is less than that with EQA and in fact close to that with EDF. EDF results in a delay jitter; however, both EQA and SWIM can achieve zero delay jitter.

In other words, there is a tradeoff of delay-jitter vs. the number of bursts. By scarifying the burst overhead, SWIM can reduce the delay jitter and vice versa. Without the delay jitter constraint, SWIM potentially behaves similar to EDF.

In addition, in terms of complexity trade-off (See also III.B), the complexity of EQA is quite straightforward – $O(n)$. With sorting, EDF complexity is $O(n\log n)$ and with swapping process, SWIM requires $n$ more steps thus resulting in $O(n^2)$.

## VI. CONCLUSIONS

In this paper, we introduced a new algorithm for UGS scheduler for the IEEE 802.16e Mobile WiMAX networks. The algorithm tries to minimize the number of bursts and gives zero delay jitter. Compared to EQA, the number of bursts is much less. Compared to EDF, the delay jitter is zero, and the number of bursts is comparable. Although this technique has been designed for UGS service, we believe a simple extension with a polling mechanism can be used for ertPS service.

There is a tradeoff between delay jitter and the number of bursts. We showed that SWIM results in less numbers of bursts than equal allocation (EQA). SWIM's burst count is comparable to that of EDF; however, SWIM achieves zero delay jitter.

In this paper, we assumed that all slots have the same capacity. With adaptive modulation, the slot capacity for each slot may be different. We are working on an extension to handle this case. Finally, we have assumed the uplink allocation. The same algorithm can be extended for the downlink allocation with further optimization using extra information such as the actual arrivals, packet sizes, and head of line delays.

More variations of configurations, topologies, simulation scenarios, different traffic types including large number of mobile stations can be investigated in future. Moreover, with Automatic Repeat reQest (ARQ) and Hybrid ARQ features enabled, the scheduler needs to accommodate scheduling of the retransmission/feedback and the boundary of ARQ block [2, 13].

One more requirement for Mobile WiMAX scheduling (without Hybrid ARQ) is that all downlink allocations be mapped to a rectangular area in the Orthogonal Frequency Division Multiple Access (OFDMA) frame. That restriction can reduce the throughput since some space may need to be left unused to make the allocation rectangular [2, 14]. In addition, the optional packing feature can also be added. These issues need more investigation.

## REFERENCES

[1] IEEE Std 802.16-2009, "IEEE Standard for Local and metropolitan area networks: Part 16: Air Interface for Broadband Wireless Access Systems," May 2009, 2094 pp.
[2] C. So-In, R. Jain, and A. Al-Tamimi, "Scheduling in IEEE 802.16e WiMAX Networks: Key Issues and a Survey," *IEEE J. on Selected Areas in Commun.*, vol. 27, no. 2, pp. 156-171, Feb. 2009.
[3] C. So-In, R. Jain, and A. Al-Tamimi, "Capacity Evaluation for IEEE 802.16e Mobile WiMAX," *J. of Computer Systems, Networks, and Commun.*, vol. 2010, April 2010.
[4] C. Cicconetti, L. Lenzini, E. Mingozzi, and C. Eklund, "Quality of service support in IEEE 802.16 networks," *IEEE Networks*, vol. 20, no. 2, pp. 50-55, April 2006.
[5] K. Wongthavarawat and A. Ganz, "IEEE 802.16 based last mile broadband wireless military networks with quality of service support," in *Proc. Military Commun. Conf.*, 2003, vol. 2, pp. 779-784.
[6] A. Sayenko, O. Alanen, and T. Hamalainen, "Scheduling solution for the IEEE 802.16 base station," *Int. J. of Comp. and Telecommunications Networking*, vol. 52, pp. 96-115, Jan. 2008.
[7] L. Dong, R. Melhem, and D. Mosse, "Effect of scheduling jitter on end-to-end delay in TDMA protocols," in *Proc. Int. Conf. on Real-Time Computing Systems and Applications*, 2000, pp. 223-230.
[8] Y. Mansour and B. Patt-Shamir, "Jitter control in QoS networks," *IEEE/ACM Trans. on Networking*, vol. 9, no. 4, pp. 492-502, Aug. 2001.
[9] D.C. Verma, H. Zhang, and D. Ferrari, "Delay jitter control for real-time communication in a packet switching network," in *Proc. Commun. for Distributed Application and Systems*, 1991, pp. 35-43.
[10] WiMAX Forum, "WiMAX System Evaluation Methodology V2.1.," July 2008, 230 pp. URL=[http://www.wimaxforum.org/technology]
[11] C. So-In, R. Jain, and A. Al-Tamimi, "SWIM: A Scheduler for Unsolicited Grant Service (UGS) in IEEE 802.16e Mobile WiMAX Networks," in *Proc. Int. Conf. on Access Networks*, 2009, pp. 40-51.
[12] R. Jain, C. So-In, and A. Al-Tamimi, "System Level Modeling of IEEE 802.16e Mobile WiMAX Networks: Key Issues," *IEEE Wireless Comm. Mag.*, vol. 15, no. 5, pp. 73-79, Oct. 2008.
[13] C. So-In, R. Jain, and A. Al-Tamimi, "OCSA: An algorithm for burst mapping in IEEE 802.16e mobile WiMAX networks," in *Proc. the 15th Asia-Pacific Conf. on Commun.*, 2009, pp. 52-58.
[14] A. Sayenko, O. Alanen, and T. Hamalainen, "ARQ aware scheduling for the IEEE 802.16 base station," in *Proc. IEEE Computer Commun. Conf.*, 2008, pp. 2667-2673.

**Chakchai So-In** received his BEng and MEng from Kasetsart University, Thailand in 1999 and 2001. He also received his MS and PhD from Washington University in St. Louis in 2006 and 2010. All are in Computer Engineering. He is currently a faculty member at the Department of Computer Science, Khon Kaen University, Thailand. In 2003, he was an Intern in a CNAP at NTU and obtained CCNP and CCDP certifications. He was Interns at Cisco Systems, WiMAX Forums and Bell Labs during summer 2006, 2008 and 2010, respectively. His research interests include architectures for future wireless networks; congestion controls; protocols to support network and transport mobility, multihoming, and privacy; and quality of service in broadband wireless access networks.

**Raj Jain** is a Professor of Computer Science and Engineering at Washington University in St. Louis. He was a Senior Consulting Engineer at Digital Equipment Corporation in Littleton, Mass and then a professor of Computer and Information Sciences at Ohio State University in Columbus, Ohio. Dr. Jain is a Fellow of IEEE, a Fellow of ACM and ranks among the top 50 in Citeseer's list of Most Cited Authors in Computer Science. He is the author of "Art of Computer Systems Performance Analysis," which won the 1991 "Best-Advanced How-to Book, Systems" award from Computer Press Association. His fourth book entitled "High-Performance TCP/IP: Concepts, Issues, and Solutions," was published by Prentice Hall in November 2003. He is also a winner of ACM SIGCOMM Test of Time award.

**Abdel Karim Al Tamimi** is currently a PhD candidate in Computer Engineering at Washington University in St. Louis, MO. Abdel Karim received his BA degree in computer engineering from Yarmouk University, Jordan. He has been awarded a full scholarship to peruse his Master and PhD degrees in computer engineering at Washington University in St. Louis, where he obtained his master degree [2007].